\documentclass{article}
\usepackage{psfrag,epsf}
\usepackage{enumerate}
\usepackage[utf8]{inputenc} 
\usepackage[T1]{fontenc}    
\usepackage{hyperref}       
\usepackage{url}            
\usepackage{booktabs}       
\usepackage{amsfonts}       
\usepackage{nicefrac}       
\usepackage{microtype}      
\usepackage{float}          
\usepackage{natbib}         
\usepackage{bm}             
\usepackage{amssymb}        
\usepackage{amsmath}        
\usepackage{graphicx}       
\usepackage{comment}
\usepackage{lscape}
\usepackage[font=footnotesize,labelfont=bf]{caption}
\usepackage{subcaption}
\usepackage{enumitem}
\usepackage{authblk}
\usepackage{dirtytalk}      
\usepackage[bb=dsserif]{mathalpha}

\newcommand{\blind}{0}

\begin{document}

\def\spacingset#1{\renewcommand{\baselinestretch}%
{#1}\small\normalsize} \spacingset{1}


\if0\blind
{
  \title{\bf Detecting consumers' financial vulnerability using Open Banking data: evidence from UK payday loans}

   \author[1]{Victor Medina-Olivares\footnote{victor.medina@ed.ac.uk}}
    \author[1]{Raffaella Calabrese}
    \affil[1]{University of Edinburgh Business School, 29 Buccleuch Place, Edinburgh EH8 9JS, UK.}
    \date{ }
  \maketitle
} \fi

\if1\blind
{
  \bigskip
  \bigskip
  \bigskip
  \begin{center}
    {\LARGE\bf Detecting consumers' financial vulnerability using Open Banking data: evidence from UK payday loans}
\end{center}
  \medskip
} \fi

\bigskip
\begin{abstract}
This paper examines whether repeated payday loan use, commonly known as the debt trap, harms borrowers' financial wellbeing. Using Open Banking data from 1,815 UK borrowers observed between 2017 and 2018, we model borrowing intensity using a two-state hidden Markov model (HMM). The HMM outperforms single-regime alternatives and identifies two distinct borrowing patterns: occasional (low-intensity) and persistent (high-intensity) use. Each regime exhibits a characteristic relationship between borrowing intensity and wider transaction behaviour. We translate the decoded state sequence into a practical monitoring rule based on sustained high-intensity exposure. Defining a trigger event as 12 consecutive weeks in the high-intensity regime, we find that 36.4\% of borrowers experience at least one such event. Among those who do, high-intensity weeks represent 17.8\% of all borrower-week observations on average. Together, these results provide evidence for a persistent high-intensity borrowing pattern and demonstrate that it can serve as a simple, interpretable rule for monitoring prolonged reliance on payday loans.

\end{abstract}

\noindent%
\vfill
{\it Acknowledgements: } Raffaella Calabrese acknowledges the support of the UK ESRC Project code ES/W010259/1
\spacingset{1.5}

\section{Introduction}

A payday loan is a short-term loan typically taken for a small amount. According to \citet{Gathergood2019}, the United Kingdom has the world's second-largest payday loan market after the United States. In 2016/7, the UK payday loan market was valued at approximately Â£220 million \citep{BBC2017}. However, due to stricter regulations, the industry has experienced a significant decline from its Â£2.5 billion valuation in 2013. Payday loan borrowers often face difficulties repaying the total balance and have to renew their loans upon maturity, incurring high fees that exacerbate their financial conditions \citep{carter2022time}. 

This repeated usage of payday loans over an extended period is commonly referred to as the debt trap \citep{CFPB2016}. To address this issue, regulators have implemented a range of policy interventions. In the United States, these include payday loan bans, mandatory cooling-off periods between consecutive loans, and restrictions on rollovers or the number of loans permitted annually \citep{Desai2017}. In the United Kingdom, the Financial Conduct Authority introduced a price cap limiting interest and fees to 0.8\% per day of the borrowed amount. Implemented in 2015, this cap reduced by 45\% the number of borrowers seeking advice on payday loan-related problems \citep{CA2016}. It also triggered a surge in customer compensation claims, contributing to the collapse of Wonga, formerly the UK's largest payday lender, in 2018.

The debt trap hypothesis suggests that payday lending exacerbates borrowers' financial vulnerability. Financial vulnerability is a broad concept encompassing various indicators of financial distress, including missed essential payments, credit delinquencies, and defaults. However, empirical evidence on the relationship between payday lending and financial vulnerability remains mixed, particularly across studies conducted in the US and the UK. Most of the existing literature focuses on the US context. For instance, \citet{Melzer2011} finds that payday loans reduce borrowers' ability to meet essential expenses such as rent, mortgage, and utility payments. Furthermore, low-income borrowers who rely on payday loans tend to exhibit higher rates of child support delinquency, increased reliance on food assistance, and greater overall financial hardship, often prioritising payday loan repayment over other obligations \citep{Melzer2018}.

A key mechanism linking payday loans to financial vulnerability is the structure of the loans themselves. Payday loans are typically characterised by high fees, short repayment periods, and lump-sum repayment requirements. For borrowers with limited liquidity, repaying the full amount on the due date often necessitates taking out a new loan, thereby initiating a cycle of repeated borrowing. Over time, this cycle increases the borrower's debt burden, reduces disposable income, and limits their ability to absorb financial shocks. As a result, individuals become increasingly exposed to financial stress, reinforcing and deepening their vulnerability.

Moreover, financial vulnerability both contributes to and is reinforced by payday loan usage, creating a feedback loop. Individuals facing income instability, lack of savings, or restricted access to mainstream credit are more likely to turn to payday loans as a coping mechanism. However, reliance on these high-cost credit products can worsen their financial condition by increasing repayment obligations and reducing financial flexibility. This dynamic illustrates how payday lending may not only be a symptom of financial vulnerability but also a driver of its persistence, trapping borrowers in a cycle of dependency and limited financial resilience.

Nevertheless, \citet{Desai2017} report insignificant changes in delinquencies following the introduction of fee caps or payday loan bans in US states such as Oregon, North Carolina, and Georgia. Interestingly, survey respondents in Oregon were more likely to report worsening financial conditions compared to respondents in Washington State, where no such ban was implemented \citep{Zinman2010}. In contrast, \citet{Morgan2012} find that Chapter 13 bankruptcy filings declined significantly following payday loan bans. In the UK context, \citet{Gathergood2019} analyse nearly the entire population of payday loans issued between 2012 and 2013 and show that, while these loans provide short-term liquidity, they also foster long-term credit dependency, leading to increased defaults and overdraft limit breaches.

To further investigate the relationship between payday lending and financial vulnerability, this study draws on a novel data source provided by an Open Banking (OB) platform. OB represents a financial ecosystem that enables secure, user-consented access to transaction-level banking data. This transformation began with the introduction of the General Data Protection Regulation (GDPR) in April 2016, which established that individuals, rather than financial institutions, own their personal data and control its use. Subsequently, the Second Payment Services Directive (PSD2), implemented in January 2018, required financial institutions to grant authorised third-party providers access to customer data via application programming interfaces (APIs), covering both retail and corporate accounts. Open Banking has the potential to enhance consumer outcomes, foster innovation, and increase competition within financial services.

The use of OB data offers significant advantages for analysing financial vulnerability. It enables a comprehensive and dynamic assessment of individuals' financial situations over time \citep{svetlovsak2023subject}, as well as comparisons across borrowers from different financial institutions. For example, the OB platform utilised in this study aggregates data from 70 UK financial institutions, providing a detailed and holistic view of borrowers' financial behaviour. Additionally, OB data improves measurement accuracy relative to survey-based studies, which often rely on self-reported information and may be subject to recall bias or misreporting \citep{Zinman2010}.

We make three main contributions to the literature on financial vulnerability. First, we link Open Banking transaction data to the study of payday-loan dynamics. We analyse a borrower-week panel for $1{,}815$ UK payday-loan customers observed over 2017--2018, combining weekly payday-loan counts with a rich set of behavioural covariates derived from categorised transaction streams (amounts and transaction counts across expenses, income, transfers, and contributions). This setting allows us to study how borrowing intensity moves together with changes in cash flows and transaction patterns over time, which is difficult to observe using typical survey-based or lender-only data sources. Conceptually, our motivation follows the view that payday loans can provide short-term liquidity for some consumers \citep{Gathergood2019}, while other evidence suggests repeated use may harm financial well-being \citep{chen2020negative,bolen2020consumers}.

Second, we model borrowing intensity through latent regimes and quantify the value of regime structure against standard count benchmarks. Markov processes are widely used in operational research to model dynamic systems with observed states \citep{petropoulos2023operational}, whereas hidden Markov models allow the underlying states to be unobserved \citep{Elliott2014}, making them well-suited to representing latent financial conditions. We propose a two-state HMM in which weekly payday-loan counts follow a state-dependent Poisson regression. The latent states represent unobserved financial conditions, allowing both the baseline intensity and the mapping from transactions to borrowing to differ across the two regimes. We also estimate an extended specification with marginal NB2 emissions via a Gamma--Poisson mixture, which accommodates overdispersion and residual variability. In both HMM specifications, the transition matrix is assumed to be shared across borrowers (time-homogeneous latent dynamics), with borrower-specific initial state probabilities, yielding a parsimonious regime model that remains (i) interpretable and (ii) operationally meaningful. Although HMMs are often applied to single stochastic processes \citep[see, e.g.][]{quirini2014creditworthiness}, our setting is a longitudinal borrower panel in which the latent regime process is inferred jointly across individuals.

To quantify what is gained from modelling latent regimes, we benchmark the HMMs against three single-regime Poisson models of increasing flexibility: (i) a GLM, (ii) a GAM with a common smooth time effect, and (iii) a GAMM that additionally includes borrower-specific random intercepts. This progression separates the contribution of shared non-linear time variation and persistent between-borrower heterogeneity in a single-regime setting. The comparison with the HMMs then assesses whether allowing for state dependence and regime switching yields additional gains. Empirically, both HMM specifications outperform the single-regime benchmarks in out-of-sample performance, and the Poisson and NB2 HMM perform very similarly, showing that the predictive gains are primarily driven by the latent-regime structure rather than by overdispersion alone.

Third, we translate the inferred regimes into a policy-relevant monitoring rule based on prolonged exposure to the high-intensity regime (State~2). Using the decoded state sequence (Viterbi path) \citep{viterbi1967error}, we define a borrower-level event as the first occurrence of $H$ consecutive weeks in State~2, with $H=12$ (approximately three months) to mirror common default horizons in credit risk \citep{malik2010modelling,basel2004international}. This yields two complementary portfolio-level summaries. First, by the end of the observation window, 36.4\% of borrowers have triggered at least one such prolonged exposure event. Second, the weekly share of borrower-week observations that are in State 2 after the borrower has already completed a 12-week spell in State 2 averages 17.8\% over the study period. We interpret this as an average default-rate analogue.

The remainder of the paper is organised as follows. Section~\ref{sec:data} describes the Open Banking dataset and the borrower-week panel. Section~\ref{sec:meth} introduces the modelling framework, starting from the Poisson HMM and then extending the emission model to a marginal NB2 specification to accommodate overdispersion. Section~\ref{sec:emp_res} reports the empirical analysis. Section~\ref{sec:concl} concludes.

\section{Data description} \label{sec:data}
A UK Open Banking platform provided anonymised transaction data for $1{,}815$ UK payday-loan customers observed over 2017--2018, comprising $126{,}188$ raw transactions across approximately 70 UK financial institutions, offering a detailed view of borrowers' financial behaviour from linked Open Banking accounts. We construct a borrower-week panel by aggregating transactions to a weekly frequency and aligning all individuals on a common weekly time index. On average, each borrower receives around eight payday loans and is observed for roughly 80 weeks, which is comparable to other datasets used in related work \citep{lawrence2008comparative,bertrand2011information}.

All transactions are categorised using the platform's Open Banking categorisation scheme. There are four broad groups covering 14 transaction categories: (i) \emph{expenses} (basic, discretionary, luxury), (ii) \emph{income} (recurrent, non-recurrent), (iii) \emph{transfers} (basic, discretionary, recurrent, non-recurrent, luxury, and other), and (iv) \emph{contributions} (pension, saving, and investment). From these categories, we derive weekly covariates capturing both the \emph{total amount} and the \emph{number of transactions} in each category, providing 28 covariates in total (14 amounts and 14 counts). The outcome variable is the weekly number of payday loans granted to each borrower.

Table~\ref{tab:desc_new} summarises the weekly covariates using borrower-weeks and reports means, standard deviations, and medians. Discretionary expenses are larger and more frequent than basic expenses (amount mean \pounds611 vs \pounds321; transaction-count mean 12.2 vs 8.56). Luxury expenses are smaller (\pounds131) and occur less often (3.48 transactions on average), with sizeable variation in amounts (SD \pounds86.2).

Income amounts are large in both categories. Non-recurrent income typically appears in about four transactions per active week (median 4.15) with moderate variation in amount (mean \pounds782, SD \pounds192). Recurrent income, in contrast, is much less frequent (mean 1.38 transactions; median 1.38) and much more variable in amount (mean \pounds853, SD \pounds375). For recurrent income, the mean exceeds the median (\pounds853 vs \pounds707), indicating a right-skewed distribution with occasional large inflows.

Transfers differ by subtype in both scale and frequency. Basic and discretionary transfers are relatively regular at roughly 3.5-3.8 transactions per active week, with moderate amounts (means \pounds89.6 and \pounds252). Non-recurrent transfers are larger (mean \pounds332; median \pounds317) but occur less often (mean and median 1.95 transactions). Several categories are close to one weekly transaction, including recurrent transfers (median 1.07) and ``other'' transfers (median 1.17), yet their amounts can be highly dispersed, especially recurrent transfers (mean \pounds107, SD \pounds148) and other transfers (mean \pounds181, SD \pounds241). Contributions are similarly sparse in frequency (medians 1.11-1.18) and strongly right-skewed in amounts, most notably saving contributions where the mean (\pounds195) is far above the median (\pounds54.2) and the SD is very large (\pounds647).
\begin{table}[hbt]
    \centering
    \footnotesize
    \tabcolsep=0.11cm
    \begin{tabular}{l rrrrrr}
        \toprule
         Category & \multicolumn{3}{c}{Amount (\pounds)} & \multicolumn{3}{c}{Transactions} \\
         & Mean & Median & SD & Mean & Median & SD \\
         \midrule
          Basic expenses & 321.31 & 259.16 & 115.77 & 8.56 & 8.26 & 1.51 \\ 
  Discretionary expenses & 610.54 & 557.33 & 171.93 & 12.18 & 11.62 & 1.56 \\ 
  Luxury expenses & 131.07 & 108.62 & 86.20 & 3.48 & 3.44 & 0.38 \\ 
  Recurrent income & 852.52 & 706.58 & 375.28 & 1.38 & 1.38 & 0.07 \\ 
  Non-recurrent income & 781.52 & 716.48 & 192.18 & 4.18 & 4.15 & 0.25 \\ 
  Basic transfers & 89.62 & 87.14 & 15.41 & 3.52 & 3.52 & 0.30 \\ 
  Discretionary transfers & 252.36 & 237.72 & 56.89 & 3.83 & 3.83 & 0.27 \\ 
  Recurrent transfers & 107.39 & 50.12 & 148.23 & 1.09 & 1.07 & 0.11 \\ 
  Non-recurrent transfers & 331.55 & 316.86 & 95.70 & 1.95 & 1.95 & 0.11 \\ 
  Luxury transfers & 180.60 & 145.04 & 109.11 & 2.03 & 1.87 & 0.61 \\ 
  Other transfers & 180.79 & 141.15 & 240.64 & 1.24 & 1.17 & 0.25 \\ 
  Pension contributions & 157.02 & 65.54 & 317.22 & 1.23 & 1.15 & 0.25 \\ 
  Saving contributions & 194.85 & 54.24 & 646.61 & 1.26 & 1.18 & 0.26 \\ 
  Investment contributions & 147.92 & 85.71 & 271.02 & 1.12 & 1.11 & 0.15 \\ 
  \bottomrule

    \end{tabular}
    \caption{Mean, median and standard deviation (SD) of weekly transaction volume and frequency across 14 categories.}
    \label{tab:desc_new}
\end{table}

\section{Methodology}\label{sec:meth}
Let $Y_{it}\in\{0,1,2,\ldots\}$ be the number of payday loans granted to borrower $i$ ($i=1,\ldots,N$) in week $t$ ($t=1,\ldots,T_i$), and let $\bm x_{it}\in\mathbb{R}^K$ denote the associated (time-varying) covariates. We aim to characterise borrowing intensity over time and relate it to an underlying notion of \emph{financial vulnerability}. Because the available data do not contain a direct label for vulnerability, we adopt a data-driven approach and represent vulnerability through a latent, discrete state process.

A natural starting point for count data is Poisson regression, and to account for between-borrower differences, one could use a Poisson regression with random effects \citep{stroup2012generalized}. However, that modelling approach does not introduce an explicit \emph{time-evolving} latent condition that can switch between periods of relative stability and distress (see Section \ref{sec:emp_res}). We therefore use Hidden Markov Models (HMMs) \citep{ghahramani2001introduction}, which couple (i) a latent Markov chain capturing unobserved regimes with (ii) a state-dependent observation model for the counts. We proceed in two steps: we first describe a parsimonious Poisson HMM, and then extend it to better capture heterogeneity and overdispersion in longitudinal data.

\subsection{Poisson HMM}\label{subsec:meth_baseline}
\paragraph{Latent dynamics.}
Define a latent state $Z_{it}\in\{1,\ldots,K_h\}$ that represents borrower $i$'s unobserved financial condition at time $t$. For each borrower $i$, $\{Z_{it}\}_{t=1}^{T_i}$ follows a first-order, time-homogeneous Markov chain with a shared transition matrix and borrower-specific initial probabilities:
\[
\Pr(Z_{i1}=k)=\pi_{ik}, \qquad
\Pr(Z_{it}=k \mid Z_{i,t-1}=j)=P_{jk}, \qquad j,k\in\{1,\ldots,K_h\}.
\]
We impose $\bm\pi_i=(\pi_{i1},\ldots,\pi_{iK_h})\in\Delta^{K_h-1}$ and each row $\bm P_{j\cdot}\in\Delta^{K_h-1}$, where $\Delta^{K_h-1}$ denotes the $(K_h-1)$-simplex.

\begin{figure}[ht]
\centering
  \includegraphics[width=0.5\textwidth]{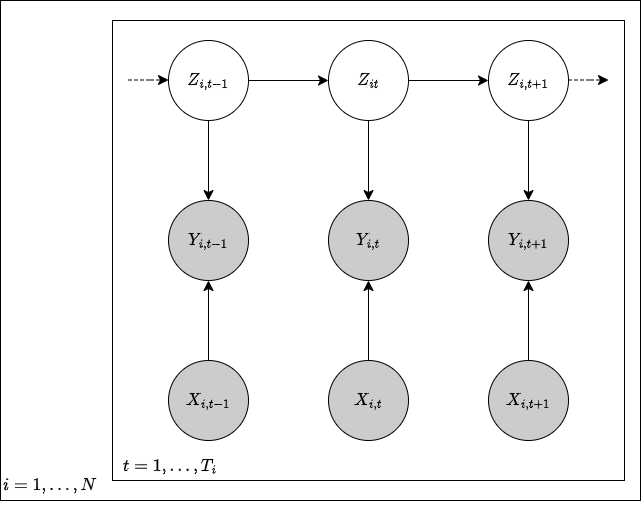}
  \caption{Graphical representation of the Poisson Hidden Markov Model.}
  \label{fig:hmm_baseline}
\end{figure}

\paragraph{State-dependent emissions.}
Conditional on the state, the weekly count follows a Poisson regression with log link and state-specific parameters:
\begin{equation}\label{eq:v0-poiss}
\begin{aligned}
Y_{it}\mid(Z_{it}=k,\bm x_{it})
&\sim \mathrm{Poisson}(\lambda_{itk}), \qquad k=1,\ldots,K_h,\\
\log \lambda_{itk}
&= \alpha_k + \bm x_{it}^\top \bm\beta_k,
\end{aligned}
\end{equation}
where $\alpha_k$ is the intercept in state $k$ and $\bm\beta_k\in\mathbb{R}^K$ is the corresponding coefficient vector. This allows covariate effects on borrowing intensity to differ across latent financial conditions. Figure~\ref{fig:hmm_baseline} illustrates the corresponding graphical model.

\paragraph{Likelihood and forward recursion.}
Let $g_k(y_{it};\bm x_{it})=p(y_{it}\mid Z_{it}=k,\bm x_{it})$ denote the Poisson pmf in \eqref{eq:v0-poiss}. For borrower $i$, define the forward variables
\[
\alpha^{(i)}_t(k)=p(y_{i1},\ldots,y_{it},Z_{it}=k), \qquad k=1,\ldots,K_h.
\]
The recursion is
\begin{align*}
\alpha^{(i)}_1(k)
&= \pi_{ik}\, g_k(y_{i1};\bm x_{i1}),\\
\alpha^{(i)}_t(k)
&= g_k(y_{it};\bm x_{it})\sum_{j=1}^{K_h}\alpha^{(i)}_{t-1}(j)\,P_{jk},\qquad t\ge 2.
\end{align*}
The likelihood contribution of borrower $i$ is $p(\bm y_i)=\sum_{k=1}^{K_h}\alpha^{(i)}_{T_i}(k)$, where $\bm y_i=(y_{i1},\ldots,y_{iT_i})^\top$. Hence, the full likelihood is
\[
\mathcal{L}(\bm\theta\mid\bm y)=\prod_{i=1}^N p(\bm y_i),
\]
with $\bm\theta$ collecting all model parameters.

\paragraph{Identification and priors.}
To anchor state interpretation and reduce label switching, we impose an ordering constraint on the state intercepts,
$\alpha_1<\alpha_2<\cdots<\alpha_{K_h}$. We use weakly informative priors
\[
\alpha_k \sim \mathcal{N}(0,3^2), \qquad
\bm\beta_k \sim \mathcal{N}(\bm 0,2^2\bm I_K),
\]
and a uniform Dirichlet prior for the borrower-specific initial distributions,
$\bm\pi_i \sim \mathrm{Dirichlet}(\mathbb{1}_{K_h})$, $i=1,\ldots,N$.

\subsection{Heterogeneity and overdispersion extension (NB2 HMM)}\label{subsec:meth_hetero}
The Poisson HMM provides a transparent link between latent financial conditions and borrowing intensity. In practice, however, longitudinal data typically exhibit (i) substantial heterogeneity in borrowing behaviour and (ii) overdispersion relative to the Poisson model. If left unmodelled, this extra variability may lead the HMM to explain residual noise through overly frequent state switching, blurring the intended interpretation of the latent process. We therefore extend the observation model to accommodate extra-Poisson variation, keeping the HMM structure and forward recursion described above.

\paragraph{Latent dynamics.}
We keep the same latent dynamics as in the Poisson HMM, i.e.\ a borrower-specific initial distribution $\bm\pi_i$ and a shared transition matrix $\bm P$:
\[
\Pr(Z_{i1}=k)=\pi_{ik}, \qquad 
\Pr(Z_{it}=k\mid Z_{i,t-1}=j)=P_{jk}, \qquad j,k\in\{1,\ldots,K_h\}.
\]
For estimation we work with unconstrained logits $\bm\eta_{\pi,i}$ and $\bm\eta_{P,j\cdot}$ and recover simplex-valued parameters via
\[
\bm\pi_i=\mathrm{softmax}(\bm\eta_{\pi,i}),\qquad
\bm P_{j\cdot}=\mathrm{softmax}(\bm\eta_{P,j\cdot}).
\]

\begin{figure}[ht]
\centering
  \includegraphics[width=0.5\textwidth]{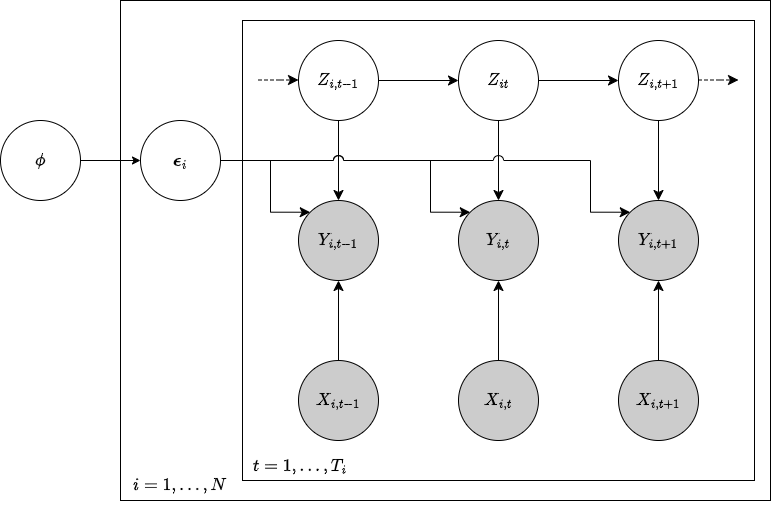}
  \caption{Graphical representation of the Negative Binomial type 2 Hidden Markov Model.}
  \label{fig:hmm_nb2}
\end{figure}

\paragraph{Emissions via a marginal Negative Binomial model.}
To capture overdispersion and residual heterogeneity without sampling explicit random effects (which is computationally demanding), we use a Gamma--Poisson mixture. Specifically, conditional on state $k$, we begin with
\begin{align}
Y_{it}\mid(Z_{it}=k,\varepsilon_{it})
&\sim \mathrm{Poisson}\!\big(\varepsilon_{it}\mu_{itk}\big),
\qquad
\log \mu_{itk}=\alpha_k+\bm x_{it}^\top\bm\beta_k,
\label{eq:poiss-frailty_rewrite}
\end{align}
and assume an observation-level Gamma frailty~\citep{cameron2013regression}
\[
\varepsilon_{it}\overset{\text{i.i.d.}}{\sim}\mathrm{Gamma}(\phi,\phi),
\quad\text{(shape $\phi>0$, rate $\phi$; mean $1$, var $1/\phi$).}
\]
Integrating out $\varepsilon_{it}$ gives the Negative Binomial type 2 (NB2) emission model
\begin{align}
Y_{it}\mid(Z_{it}=k)
&\sim \mathrm{NegBin}_2\!\big(\mu_{itk},\phi\big),
\label{eq:nb2_rewrite}
\end{align}
with $\mathbb{E}[Y_{it}\mid Z_{it}=k]=\mu_{itk}$ and
$\mathrm{Var}(Y_{it}\mid Z_{it}=k)=\mu_{itk}+\mu_{itk}^2/\phi$.
This extension preserves the conditional independence of $Y_{it}$ given $Z_{it}$ and therefore retains the standard forward recursion. Figure~\ref{fig:hmm_nb2} shows the graphical model for this specification.

\paragraph{Likelihood and computation.}
Let $g_k(y_{it};\bm x_{it})=p(y_{it}\mid Z_{it}=k,\bm x_{it})$ denote the NB2 pmf implied by \eqref{eq:nb2_rewrite}, evaluated at $\mu_{itk}=\exp(\alpha_k+\bm x_{it}^\top\bm\beta_k)$. The forward variables and recursion are identical to the Poisson model, with $g_k(\cdot)$ replaced by the NB2 emission probability. The full likelihood remains $\prod_{i=1}^N p(\bm y_i)$ and is evaluated on the log scale for numerical stability.

The Gamma--Poisson construction in \eqref{eq:poiss-frailty_rewrite}--\eqref{eq:nb2_rewrite} can be interpreted as a Poisson mixed model in which a multiplicative random effect has been marginalised out. This yields a closed-form Negative Binomial likelihood, preserves the standard HMM recursion, and, in our case, empirically improved MCMC efficiency relative to sampling large numbers of explicit subject-level random effects within the state-space. 

\paragraph{Identification and priors.}
We again order the state intercepts $\alpha_1<\cdots<\alpha_{K_h}$ to mitigate label switching and place weakly informative Gaussian priors on $\alpha_k$ and $\bm\beta_k$, Gaussian priors on the unconstrained logits $\bm\eta_{\pi,i}$ and $\bm\eta_{P,j\cdot}$, and a Gamma prior on the overdispersion $\phi$.

\paragraph{Estimation and posterior predictive simulation.}
All models are implemented in \emph{Stan}~\citep{carpenter2017stan} and fitted using MCMC\footnote{The three benchmark models described in Section~\ref{sec:emp_res} were also implemented in Stan.}.

Posterior predictive quantities can be obtained in two ways. First, for in-sample replication, we condition on a decoded state sequence (the Viterbi path~\citep{viterbi1967error}) and simulate
\[
Y^{\mathrm{rep}}_{it}\mid \hat Z_{it}=k \sim \mathrm{NegBin}_2\!\big(\exp(\alpha_k+\bm x_{it}^\top\bm\beta_k),\phi\big)
\]
for each posterior draw. Second, for forecasting on new sequences, we simulate the latent chain forward. That is, at the start of a sequence we draw $Z_{i1}\sim\mathrm{Categorical}(\mathbb{1}_2/2)$ for unseen borrowers, then for $t\ge 2$ we draw $Z_{it}\sim\mathrm{Categorical}(\bm P_{Z_{i,t-1}})$. Conditional on the simulated states, we then sample counts from the emission model as above.

\section{Empirical results} \label{sec:emp_res}
This section evaluates and compares predictive performance and then studies the estimated latent structure to assess whether the evidence is consistent with a debt-trap mechanism. We proceed in four steps. First, we compare predictive accuracy across competing models. Second, we characterise the latent regimes using the state-dependent emission coefficients. Third, we analyse transition dynamics to quantify persistence and switching between regimes. Fourth, we construct a policy-oriented measure based on sustained exposure to the high-intensity regime.

\subsection{Benchmarks and predictive performance comparison}\label{subsec:emp_perf}
We assess whether allowing for latent regimes improves predictive performance relative to several single-regime benchmarks. This comparison provides empirical motivation for the HMM specifications described in Section~\ref{sec:meth}.

\paragraph{Benchmarks.}
We consider three Poisson count models of increasing flexibility as benchmarks. The first is a single-regime Poisson regression (hereafter \emph{Poisson GLM}) with log link,
\[
Y_{it}\sim\text{Poisson}(\lambda_{it}), 
\qquad 
\log \lambda_{it}=\alpha+\bm x_{it}^\top\bm\beta,
\]
which provides a natural starting point for weekly payday loan counts.

The second benchmark augments the GLM with a flexible calendar-time component through a spline basis under the framework of generalised additive models (hereafter \emph{Poisson GAM}) \citep{hastie2017generalized,wood2017generalized}. Let $\bm b(t)\in\mathbb{R}^{B}$ denote spline basis functions evaluated at calendar week $t$ and let $\bm\gamma\in\mathbb{R}^{B}$ be the corresponding coefficients. The model is
\[
Y_{it}\sim\text{Poisson}(\lambda_{it}), 
\qquad 
\log \lambda_{it}=\alpha+\bm x_{it}^\top\bm\beta+\bm b(t)^\top\bm\gamma,
\]
which captures common non-linear temporal patterns shared across borrowers, such as seasonality or gradual changes in market conditions.

The third benchmark further adds borrower-specific random intercepts under the framework of generalised additive mixed models (hereafter \emph{Poisson GAMM}) \citep{wood2017generalized}. Let $u_i$ be a borrower-level random intercept. The model is
\[
Y_{it}\sim\text{Poisson}(\lambda_{it}), 
\qquad 
\log \lambda_{it}=\alpha+\bm x_{it}^\top\bm\beta+\bm b(t)^\top\bm\gamma+u_i,
\qquad
u_i\sim\mathcal{N}(0,\sigma_u^2),
\]
which captures temporal patterns via a spline basis and persistent differences in baseline borrowing propensity across borrowers through the random effects $u_i$.

These benchmarks add flexibility step by step to capture two key patterns in the data, common non-linear time effects and persistent differences across borrowers. Comparing the HMMs to these approaches assesses whether introducing latent regimes and state dependence yields additional predictive gains in this setting.

\paragraph{Predictive performance comparison.}
We perform a borrower-level train--test split to avoid leakage across time within the same individual. Specifically, we randomly sample 80\% of borrowers into the training set and assign the remaining 20\% to the test set. All models are estimated on the training set and evaluated on both sets using mean per-borrower mean squared error (MSE) and mean absolute error (MAE), computed by first averaging errors within each borrower and then averaging across borrowers. For the GAMM benchmark, which allows for heterogeneity, predictions for borrowers unseen during estimation marginalise over the random intercept distribution \citep{mcculloch2004generalized}. For the HMMs (Poisson and NB2), test-set predictions are obtained from the posterior predictive distribution by simulating the latent chain using the estimated transition matrix and initial distribution, and then sampling outcomes from the state-dependent emission distribution.

Table~\ref{tab:perf} reports predictive performance for all five models. For this dataset, the two HMM specifications tend to deliver better predictive accuracy than the three single-regime benchmarks, most notably in terms of MAE on the test set. The Poisson HMM attains the lowest out-of-sample errors among the models considered, with the NB2 HMM performing very similarly. Relative to the GLM and the GAM benchmarks, these gains suggest that flexible time variation alone does not capture all of the observed heterogeneity in payday-loan intensity. The comparison with the GAMM benchmark further indicates that allowing only for persistent borrower-level differences leaves residual structure in the data. Overall, the results are consistent with the view that, in this setting, modelling payday-loan borrowing through latent regimes and state dependence can be a useful and empirically relevant extension.

\begin{table}[hbt]
\centering

\begin{tabular}{lrrrr}
\toprule
\multicolumn{1}{c}{ } & \multicolumn{2}{c}{Training} & \multicolumn{2}{c}{Test} \\
\cmidrule(l{3pt}r{3pt}){2-3} \cmidrule(l{3pt}r{3pt}){4-5}
Model & MSE & MAE & MSE & MAE\\
\midrule
Poisson GLM & 0.1627 & 0.2152 & 0.1941 & 0.2287\\
Poisson GAM & 0.1625 & 0.2163 & 0.1937 & 0.2278\\
Poisson GAMM & 0.1403 & 0.1934 & 0.1944 & 0.2313\\
Poisson HMM & \textbf{0.1375} & \textbf{0.1757} & \textbf{0.1888} & \textbf{0.2227}\\
NB2 HMM & 0.1390 & 0.1781 & 0.1904 & 0.2311\\
\bottomrule
\end{tabular}
\caption{Predictive performance (mean per-borrower errors) on training and test sets.}
\label{tab:perf}
\end{table}

\subsection{Regime characterisation from emission coefficients}\label{subsec:emp_emissions}

We characterise the latent states using the state-dependent emission model of the Poisson HMM, which achieved the best predictive performance in Table~\ref{tab:perf}. The corresponding results for the NB2 HMM are qualitatively similar and are reported in Appendix~\ref{app:coeff_nb2}. Figure~\ref{fig:coeffs2_base_hmm} reports posterior intervals for the HMM coefficients and compares them with the coefficients from the best single-regime benchmark (Poisson GAMM), shown as red vertical lines. 

\begin{figure}[ht!]
\centering
  \includegraphics[width=\textwidth]{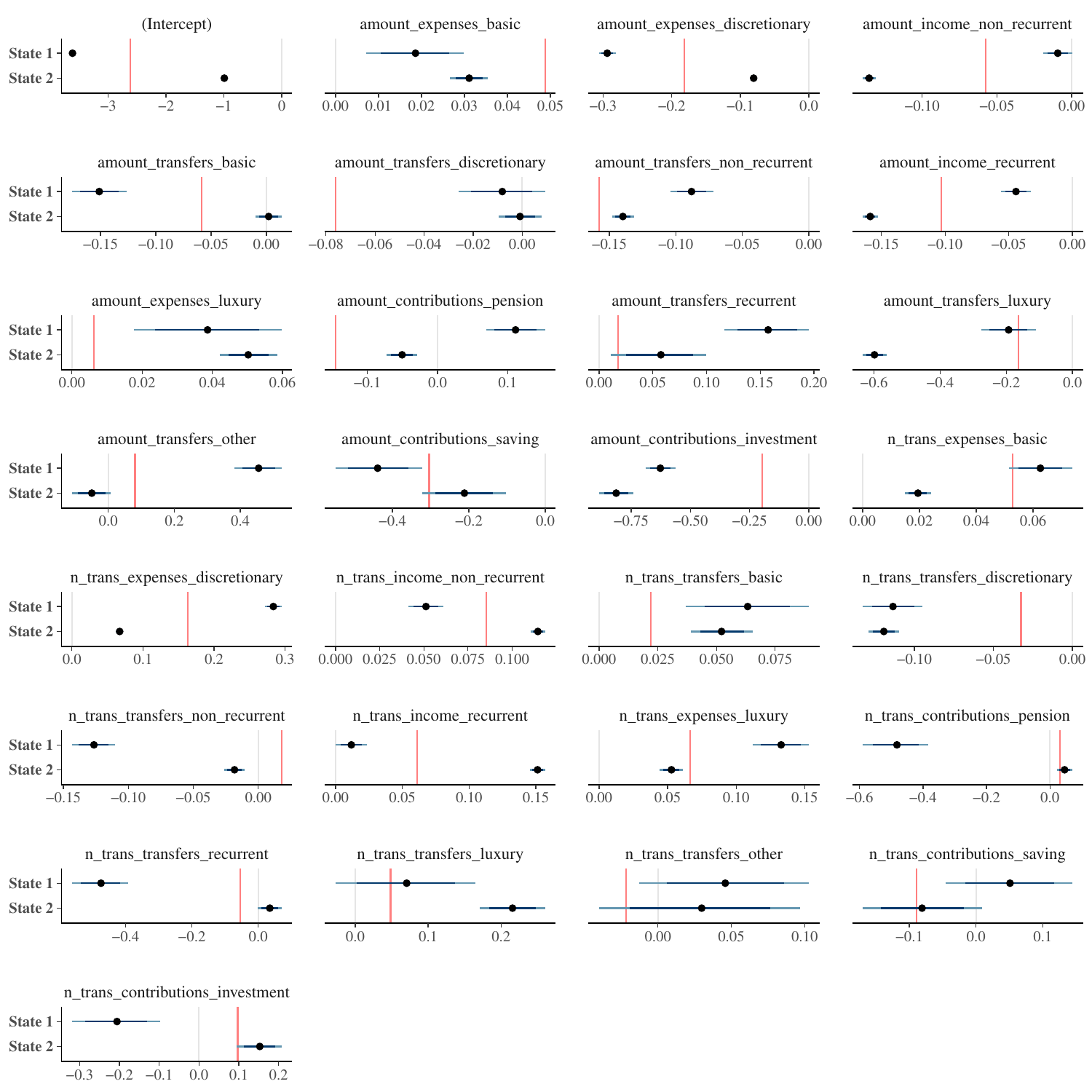}
  \caption{Posterior estimates of the coefficients for the Poisson HMM with two states. Points denote posterior medians, the inner and outer segments denote 75\% and 90\% credible intervals. Red vertical lines show coefficients from the Poisson GAMM benchmark.}
  \label{fig:coeffs2_base_hmm}
\end{figure}

The clearest separation between the two states is in the intercept. State~1 has a very low intercept ($\alpha_1\approx-3.62$), implying a baseline mean of roughly $\exp(\alpha_1)\approx 0.03$ payday loans per week. State~2 has a much higher intercept ($\alpha_2\approx-0.99$), implying $\exp(\alpha_2)\approx 0.37$ payday loans per week. This corresponds to an increase of approximately fourteen-fold in baseline intensity. We therefore refer to State~1 as a \emph{low-intensity (occasional-use) regime} and State~2 as a \emph{high-intensity (persistent-use) regime}.

Beyond the intercept, the two regimes differ in how transaction patterns map into borrowing behaviour.

First, the low-intensity regime (State~1) is more sensitive to transaction frequency, especially in spending categories. The coefficients on the number of expense transactions are substantially larger in this state than in State~2, particularly for discretionary and luxury expenses. This pattern suggests that borrowing is occasional and tends to occur in weeks with unusually high spending or more day-to-day transactions.

Second, the high-intensity regime (State~2) is more strongly related to income. In this state, both recurrent and non-recurrent income amounts have clearly negative coefficients, and they are much more negative than in State~1. This means that, even among high-use borrowers, weeks with larger inflows are associated with fewer payday loans.

At the same time, some coefficients for the number of income transactions are positive in State~2. This shows a different aspect of income dynamics. Holding the total amount constant, income arriving through more separate transactions is associated with higher borrowing intensity. An interpretation of that is, in the persistent-use regime, not only how much income arrives matters, but also how it arrives. Large inflows ease borrowing needs, but more fragmented or irregular income patterns are associated with higher reliance on payday loans.

Third, transfers and contributions help distinguish the two regimes. Some transfer categories behave very differently across states. For example, higher luxury-transfer amounts are much more strongly associated with lower payday-loan intensity in State~2 than in State~1. By contrast, ``other'' transfer amounts are positively associated with payday-loan intensity in State~1, but this relationship largely disappears in State~2.

Contribution variables also differ by regime. In particular, higher saving and investment contribution amounts are associated with lower payday-loan intensity, and this negative relationship is stronger in the persistent-use regime. One way to interpret this is that, within each regime, weeks where borrowers direct more funds to savings or investments tend to coincide with a reduced need for payday loans.

These patterns show that State~2 is not only a ``higher-borrowing'' version of State~1. The relevant categories, and the direction and strength of their associations, differ across these two regimes.

Moreover, under a single-regime specification, these patterns are largely averaged, as illustrated by the Poisson GAMM benchmark estimates (red vertical lines). In Figure~\ref{fig:coeffs2_base_hmm}, many benchmark coefficients lie between the two regime-specific estimates, and the benchmark intercept lies between $\alpha_1$ and $\alpha_2$. As a result, a single-regime model compresses distinct behavioural relationships into a single compromise pattern and does not recover the sharp separation between occasional-use and persistent-use borrowing that emerges from the hidden states.

Importantly, this averaging occurs at the level of the \emph{fixed effects}. The Poisson GAMM can still accommodate substantial heterogeneity through the borrower-specific random intercepts. Appendix~\ref{app:re_heterogeneity} shows that these random intercepts $u_i$ are in fact strongly associated with HMM regime occupancy. That is, borrowers with higher $u_i$ spend a much larger share of weeks in the high-intensity regime (State~2), with a Spearman correlation of $\rho=0.83$ and a steep transition captured by a binomial-logit model (Figure~\ref{fig:re_state2_share}). In other words, the GAMM captures persistent heterogeneity via a static borrower effect, whereas the HMM captures both heterogeneity and \emph{within-borrower switching} between low- and high-intensity regimes.

\begin{figure}[ht!]
\centering
  \includegraphics[width=0.9\textwidth]{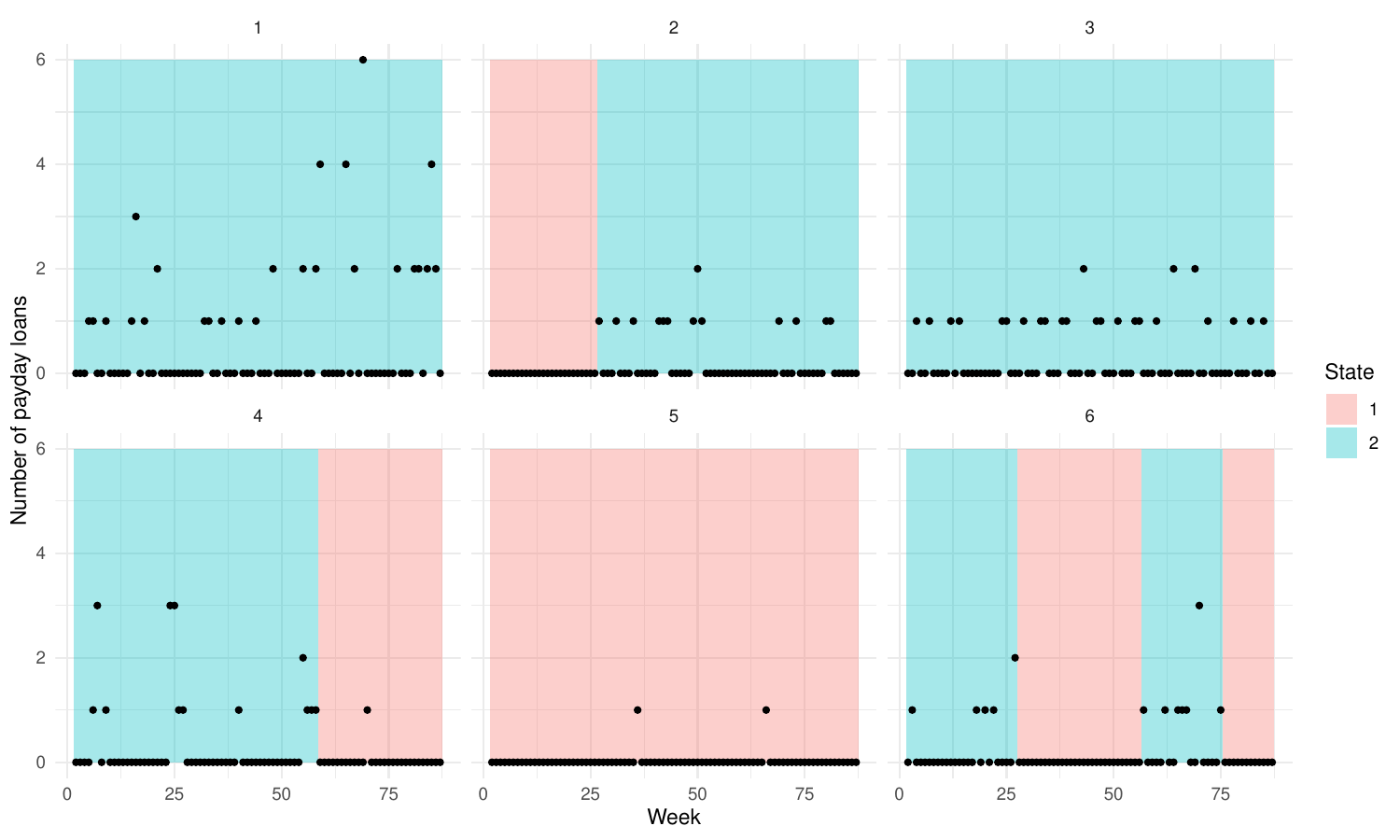}
  \caption{Evolution in time of the predictive states for the six selected borrowers. The dots represent the number of payday loans granted.}
  \label{fig:predict_states2}
\end{figure}

For interpretability, Figure~\ref{fig:predict_states2} plots the most probable state sequence for six selected borrowers together with their observed weekly number of payday loans. The figure illustrates both persistent regime membership and within-borrower switching. Borrowers 1 and 3 are classified entirely in State~2 and display sustained borrowing activity. Borrower 5 is classified almost entirely in State~1, with very limited borrowing. Borrowers 2 and 4 show clear regime changes. Borrower 2 is initially classified in State~1 with little activity and then transitions into State~2 after roughly 25 weeks, coinciding with an increase in loan activity. Borrower 4 shows the opposite pattern, spending an extended period in State~2 before transitioning into State~1 later in the study period. Finally, borrower 6 exhibits multiple switches between the two states, with alternating periods of higher and lower borrowing intensity. 

\subsection{Transition dynamics}\label{subsec:emp_transitions}
Table~\ref{tab:post_trans} reports posterior median transition probabilities and 95\% credible intervals for the Poisson HMM. Analogous results for the NB2 HMM are reported in Appendix~\ref{app:trans_nb2}. The estimated transition matrix shows that \emph{both} regimes are highly persistent. Borrowers remain in the low-intensity (occasional-use) regime with probability $P_{11}\approx 0.992$ per week and remain in the high-intensity (persistent-use) regime with probability $P_{22}\approx 0.969$ per week. These magnitudes imply long expected spell durations in each regime. Under constant transition probabilities (time-homogeneous latent dynamics), the expected duration in state $k$ is $1/(1-P_{kk})$. Using the posterior medians, the implied expected duration is approximately $1/(1-0.992)\approx 125$ weeks in State~1\footnote{This expected duration exceeds the average observed window, which reflects the very high estimated persistence and should be interpreted as high stability over the available horizon.} and $1/(1-0.969)\approx 32$ weeks in State~2. In other words, the model suggests that the high-intensity regime typically persists for several months once entered, while the low-intensity regime is even more stable.
\begin{table}[hbt]
    \centering
    \footnotesize
    \tabcolsep=0.12cm
    \begin{tabular}{l rr}
        \toprule
        & \multicolumn{2}{c}{To state} \\
        From state & State 1 & State 2 \\
        \midrule
           1 & 0.992 [0.991, 0.992] & 0.008 [0.008, 0.009] \\ 
    2 & 0.031 [0.030, 0.032] & 0.969 [0.968, 0.970] \\ 
  \bottomrule

    \end{tabular}
    \caption{Estimated transition probabilities (posterior median and 95\% credible interval).}
    \label{tab:post_trans}
\end{table}

Switching between regimes is uncommon on a week-to-week basis and is asymmetric. Entry into the high-intensity regime is rare, with $P_{12}\approx 0.008$, whereas exit from the high-intensity regime occurs more frequently, with $P_{21}\approx 0.031$, although persistence remains strong because $P_{22}$ is close to one. These estimates imply that most borrower-weeks are spent in the low-intensity regime, while a non-trivial share occur in the high-intensity regime. A useful summary is the stationary probability of State~2 for a two-state Markov chain, $\pi_2 = P_{12}/(P_{12}+P_{21})$, which yields $\pi_2 \approx 0.21$ using the posterior medians.

These transition dynamics are directly relevant to the debt-trap concept. A debt trap does not require a high-intensity regime to be absorbing. Rather, it requires that borrowers who enter a high-intensity borrowing regime tend to remain there for extended periods. The estimated value of $P_{22}$ and the implied expected duration of roughly 32 weeks are consistent with persistent episodes of elevated payday-loan intensity. Combined with the emission results in Section~\ref{subsec:emp_emissions}, the transitions support an interpretation in which a relatively small segment of borrower-weeks is characterised by persistent high-intensity borrowing, while most borrower-weeks remain in the low-intensity regime.

\subsection{Policy measure based on prolonged exposure}\label{subsec:emp_policy}
The estimated hidden state sequences allow us to translate the latent regimes into a monitoring rule based on prolonged exposure to the high-intensity regime (State~2). Let $\hat{Z}_{it}\in\{1,2\}$ denote the estimated most probable state for borrower $i$ at week $t$. We define a \emph{default} event as the first time a borrower experiences $H$ consecutive weeks in State~2, with $H=12$ in our application. This horizon corresponds to roughly three months, which is commonly used in credit risk definitions of default \citep{basel2004international}.

Formally, let $T_i^{(H)}$ be the first week such that $\hat{Z}_{i,t-H+1}=\cdots=\hat{Z}_{it}=2$. We define the indicator
\[
D_{it}=\mathbb{1}\{t\ge T_i^{(H)}\}\,\mathbb{1}\{\hat{Z}_{it}=2\},
\]
which flags borrower-week observations in which the borrower has already crossed the prolonged-exposure threshold and remains in State~2. Borrowers who never reach $H$ consecutive weeks in State~2 have $D_{it}=0$ for all $t$.

\begin{figure}[ht!]
  \begin{minipage}[c]{0.45\linewidth}
  \includegraphics[width=\linewidth]{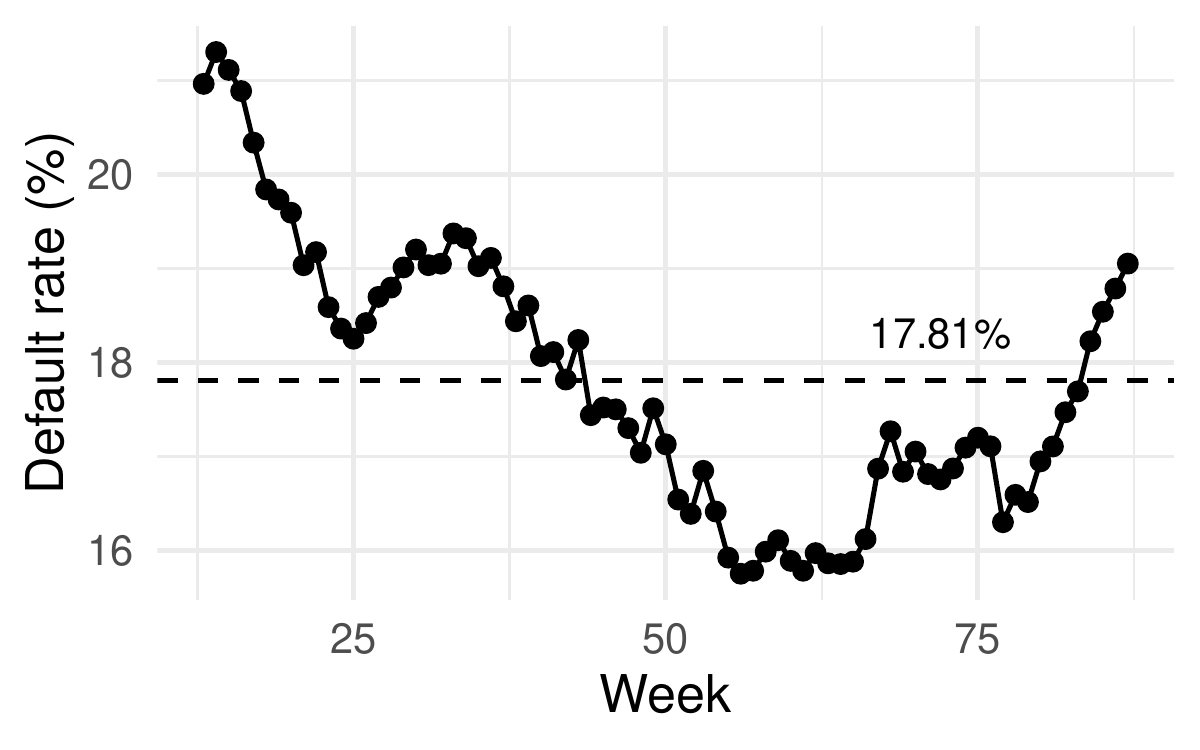}
  \caption{Average of $D_{it}$ by week, interpreted as a default-rate analogue.}
  \label{fig:df_policy2}
  \end{minipage}
  \hfill
  \begin{minipage}[c]{0.45\linewidth}
  \includegraphics[width=\linewidth]{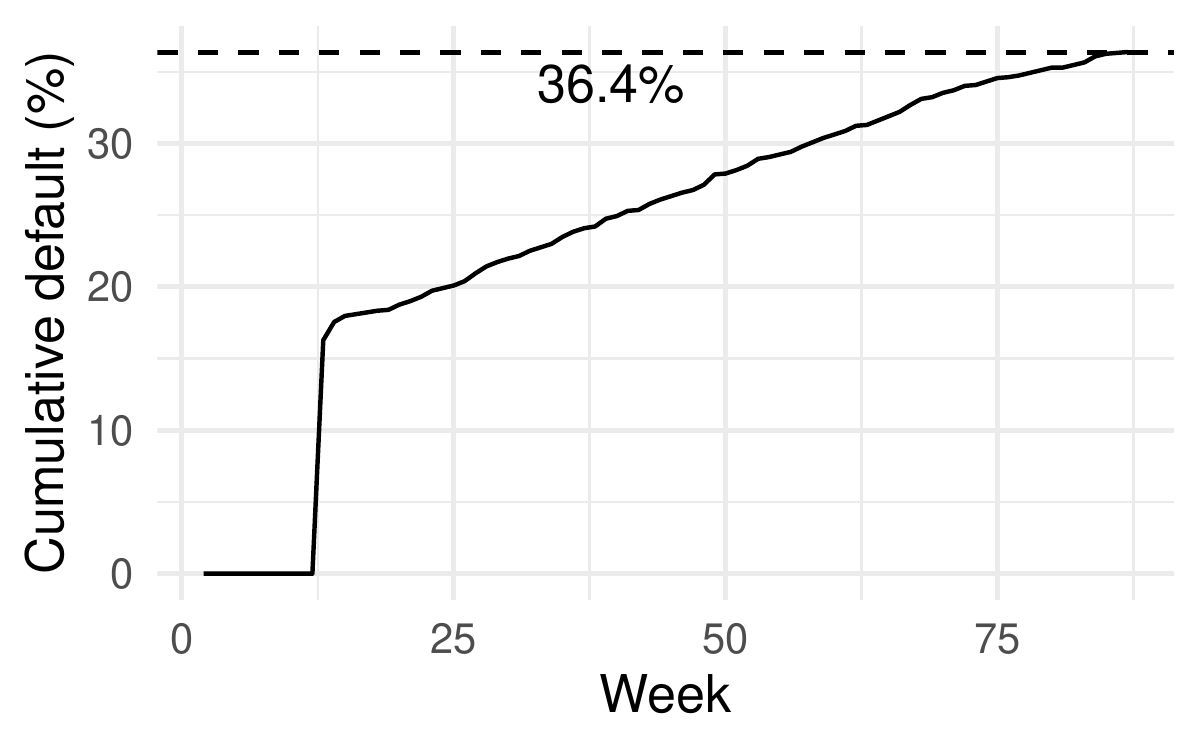}
  \caption{Cumulative share of borrowers who trigger default under the $H=12$ policy.}
  \label{fig:cum_tran}
  \end{minipage}%
\end{figure}

Figure~\ref{fig:df_policy2} reports the average of $D_{it}$ by calendar week across all observed borrower-week records. This series is a default-rate analogue because it measures, by week, the share of borrower-week observations that are in State 2 after the 12-week threshold has been reached. Over the study period, this rate averages about $17.8\%$ with relatively stable variation.

Figure~\ref{fig:cum_tran} shows a complementary borrower-level view by plotting the cumulative share of borrowers who have triggered the default event by each week. By the end of the study period, approximately $36.4\%$ of borrowers have experienced at least one event of $12$ consecutive weeks in State~2. Hence, Figures~\ref{fig:df_policy2}--\ref{fig:cum_tran} separate two perspectives: (i) how common post-threshold State~2 weeks are in the panel, and (ii) how many borrowers ever reach the threshold during the study period.

This policy complements the transition analysis in Table~\ref{tab:post_trans}. Table~\ref{tab:post_trans} summarises one-step persistence and switching probabilities, whereas Figures~\ref{fig:df_policy2}--\ref{fig:cum_tran} translate these dynamics into portfolio-level monitoring summaries constructed from the estimated state sequences. Figure~\ref{fig:cum_tran} reports the cumulative share of borrowers who have \emph{ever} triggered the event by each week, where the event time is the first week at which a run of $H$ consecutive weeks in State~2 is completed. By contrast, Figure~\ref{fig:df_policy2} reports a prevalence measure. That is, in each week, it shows the share of borrower-week observations that are both (i) \emph{post-threshold} (the borrower has already triggered the event at some earlier week) and (ii) currently in State~2. Thus, Figure~\ref{fig:cum_tran} quantifies how many borrowers cross the threshold over time, while Figure~\ref{fig:df_policy2} quantifies how common post-threshold high-intensity weeks are in the panel. The latter is a ``default rate'' analogue in the sense that it measures the proportion of the portfolio that is simultaneously beyond the prolonged exposure threshold and in the high-intensity regime in a given week.

In the context of the debt-trap hypothesis, we found that the prolonged exposure to the high-intensity regime is not rare. Approximately one-third of borrowers reach the 12-week threshold during the study period, and post-threshold weeks in State~2 account for a non-trivial share of borrower-week observations over time. Additional evidence in Appendix~\ref{app:re_heterogeneity} shows that borrowers flagged by this prolonged exposure policy concentrate in the upper tail of the Poisson GAMM random intercept distribution (Figure~\ref{fig:re_u_by_default}). Under a Gaussian mixture analysis fitted to the random intercepts, we found that 81.5\% of inferred defaults fall in the highest-intercept component, compared with 3.8\% of non-defaults (Table~\ref{tab:re_default_comp}). This agreement between the two models suggests the policy is capturing a real pattern of sustained high borrowing.

\section{Conclusions} \label{sec:concl}
This paper studies the debt-trap hypothesis in payday lending using Open Banking transaction data. We analyse a borrower-week panel of $1{,}815$ UK payday-loan customers observed over 2017--2018 and relate weekly payday-loan intensity to rich behavioural covariates capturing cash-flow and transaction patterns across expenses, income, transfers, and contributions. Because the data do not contain a direct label of financial vulnerability, we adopt a data-driven approach and represent vulnerability through latent regimes of borrowing intensity.

Our main empirical framework is a two-state hidden Markov model with state-dependent Poisson emissions. The estimated regimes admit a clear interpretation. The separation between states is most pronounced in the baseline intensity. State~1 corresponds to occasional use (low weekly intensity) and State~2 corresponds to persistent use (high weekly intensity). Beyond this level shift, the mapping from transaction behaviour to borrowing differs by regime. The low-intensity regime is more sensitive to transaction frequency in spending categories, consistent with occasional borrowing in weeks of elevated day-to-day activity. The high-intensity regime is more strongly related to income dynamics. Larger income inflows are associated with lower borrowing intensity, yet more fragmented income arrival patterns can coincide with higher borrowing, suggesting that regularity of cash flow matters in the persistent-use regime. Transfers and contributions further differentiate the regimes, with stronger negative associations between saving and investment contributions and payday-loan intensity in the high-intensity regime.

We benchmark the HMMs against three single-regime Poisson models of increasing flexibility: a GLM, a GAM with a common smooth time effect, and a GAMM that adds borrower-specific random intercepts. Both HMM specifications achieve better out-of-sample performance than these benchmarks, with the Poisson and NB2 HMM performing very similarly. This pattern indicates that the predictive gains are driven primarily by latent-regime structure and state dependence, not only by flexible time effects or persistent borrower heterogeneity. The comparison with the Poisson GAMM also clarifies the interpretation. Many GAMM fixed-effect estimates fall between the two regime-specific coefficients, consistent with a single-regime model averaging over distinct behavioural relationships. The GAMM still captures important between-borrower differences through its random intercepts, and these random effects align strongly with regime occupancy, reinforcing the interpretation that the HMM is separating a persistent high-intensity regime from more occasional borrowing.

The estimated transition dynamics show that both regimes are highly persistent. Entry into the high-intensity regime is uncommon on a week-to-week basis, and once borrowers enter State~2, they typically remain there for extended spells. This persistence provides direct support for the debt-trap mechanism, which does not require an absorbing high-intensity state but does require that high-intensity borrowing episodes tend to last for months.

We translate the inferred regimes into a monitoring rule based on prolonged exposure to the high-intensity regime. Defining a borrower-level event as the first occurrence of 12 consecutive weeks in State~2 yields two portfolio-level summaries. By the end of the observation window, 36.4\% of borrowers have triggered at least one prolonged-exposure event. The weekly prevalence of post-threshold high-intensity observations averages 17.8\% over the study period. These results indicate that sustained high-intensity borrowing is not rare in this portfolio and can be summarised in a way that is directly interpretable for monitoring purposes.

Several limitations motivate further work. The OB panel records observed behaviour over a finite period, but it does not include direct outcomes such as arrears, credit-bureau defaults, or lender profitability. A useful next step would be to link the inferred regimes to these outcomes, for example through joint longitudinal-survival models \citep{medina2025joint_spatio}. Moreover, the model uses time-homogeneous transitions that are common to all borrowers. This makes the regime dynamics easier to interpret, but it may overlook differences in switching behaviour across borrowers or over the economic cycle \citep{gu2014reduced}. Extensions could therefore allow transition probabilities to vary with borrower characteristics or macroeconomic indicators. It could also combine the regime process with additional outcome data, so that the inferred states can be related more directly to external measures of financial distress. Semi-structured multi-state models of delinquency transitions provide one possible route for doing this \citep{medina2026semi}.

Ultimately, our study sheds light on the complex relationship between financial vulnerability and payday lending. Our findings highlight the need for a more nuanced approach to responsible lending that recognises the different levels of vulnerability among borrowers. By understanding the factors contributing to the debt trap associated with payday loans, policymakers and lenders can mitigate the risks and provide better support to those who need it most.

\vspace{0.5cm}
{\noindent\it Declaration of Competing Interest:} None.

\clearpage

\bibliographystyle{apalike}

\bibliography{references}

\clearpage

{\LARGE{\begin{center}{\textbf{APPENDIX}}\end{center}}}
\appendix

\section{Posterior estimates of the coefficient NB2 HMM model} \label{app:coeff_nb2}

\begin{figure}[ht!]
\centering
  \includegraphics[width=\textwidth]{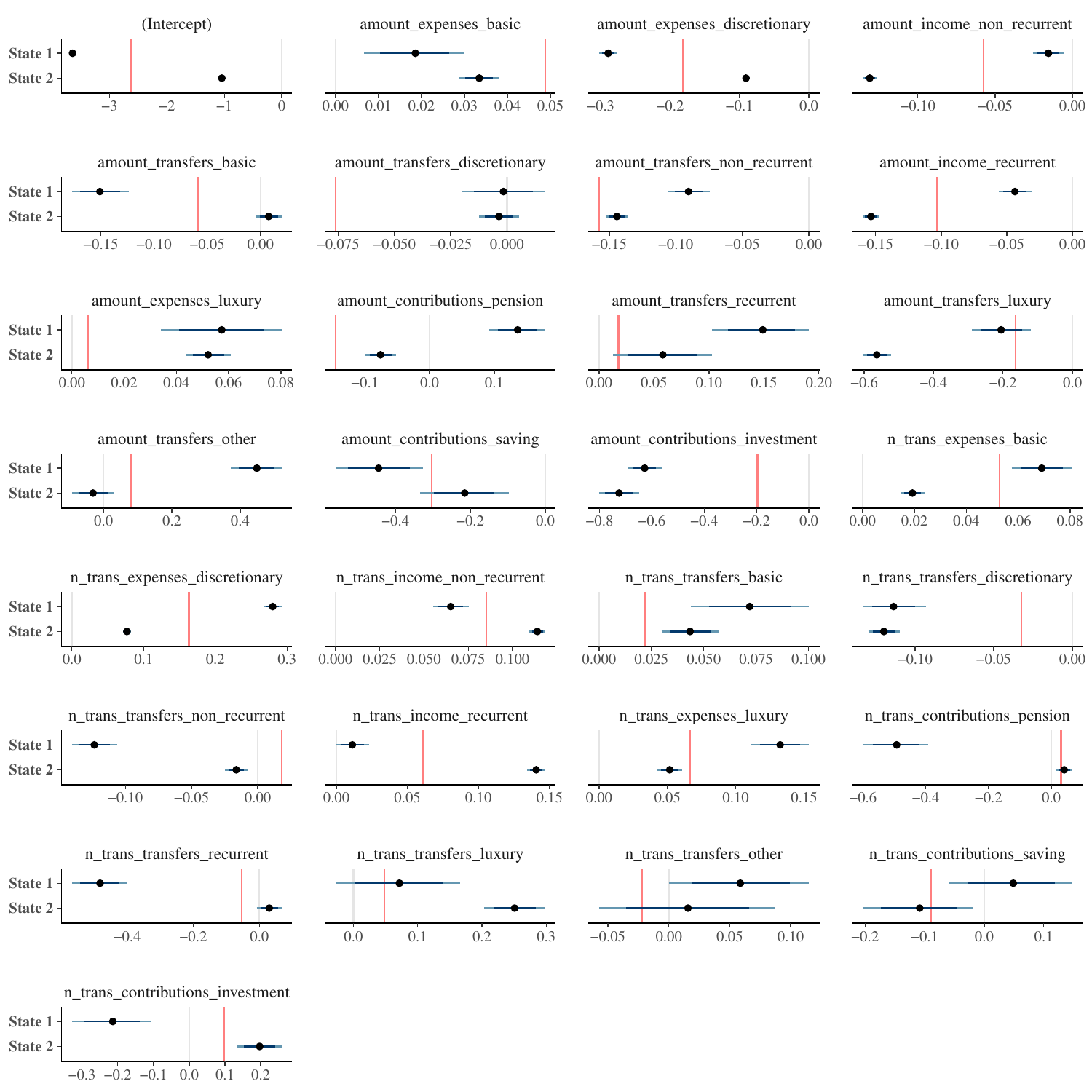}
  \caption{Posterior estimates of the coefficients for the NB2 HMM with two states. Points denote posterior medians, the inner and outer segments denote 75\% and 90\% credible intervals. Red vertical lines show coefficients from the Poisson GAMM benchmark.}
  \label{fig:coeffs3}
\end{figure}

Figure~\ref{fig:coeffs3} displays posterior medians and credible intervals for the state-dependent emission coefficients, with the Poisson GAMM benchmark overlaid for reference (red vertical lines). The qualitative conclusions are unchanged relative to the Poisson HMM. That is, the two regimes remain clearly separated, and the signs and relative magnitudes of covariate effects are broadly consistent across model specifications. Any differences mainly reflect the additional flexibility of the NB2 model in capturing overdispersion which does not reflect into the predictive performance.

\clearpage

\section{Posterior estimates of the transition probabilities NB2 HMM} \label{app:trans_nb2}

Table~\ref{tab:trans_nb2} shows posterior summaries of the transition matrix for the NB2 HMM. Each entry gives the posterior median transition probability from one state to the other, with a 95\% credible interval. We observe that the NB2 HMM yields transition dynamics that are qualitatively similar to the Poisson HMM specification.

\begin{table}[hb]
    \centering
    \footnotesize
    \tabcolsep=0.12cm
    \begin{tabular}{l rr}
        \toprule
        & \multicolumn{2}{c}{To state} \\
        From state & State 1 & State 2 \\
        \midrule
           1 & 0.992 [0.992, 0.992] & 0.008 [0.008, 0.008] \\ 
    2 & 0.027 [0.027, 0.028] & 0.973 [0.972, 0.973] \\ 
  \bottomrule

    \end{tabular}
    \caption{Estimated transition probabilities for NB2 HMM (posterior median and 95\% credible interval).}
    \label{tab:trans_nb2}
\end{table}

\section{Borrower heterogeneity and the Poisson GAMM random intercept}\label{app:re_heterogeneity}

This section studies the link between the borrower-specific random intercepts from the Poisson GAMM benchmark to the latent regimes inferred by the Poisson HMM. The Poisson GAMM introduces a time-invariant borrower effect $u_i$ that shifts the log-intensity for borrower $i$ and captures persistent heterogeneity not explained by observed covariates. The Poisson HMM, on the other hand, represents heterogeneity partly through time-varying regime membership and allows borrowers to switch between a low-intensity and a high-intensity regime.

\subsection{State-2 occupancy as a function of the random intercepts}\label{app:re_occupancy}

\begin{figure}[ht!]
\centering
  \includegraphics[width=0.9\textwidth]{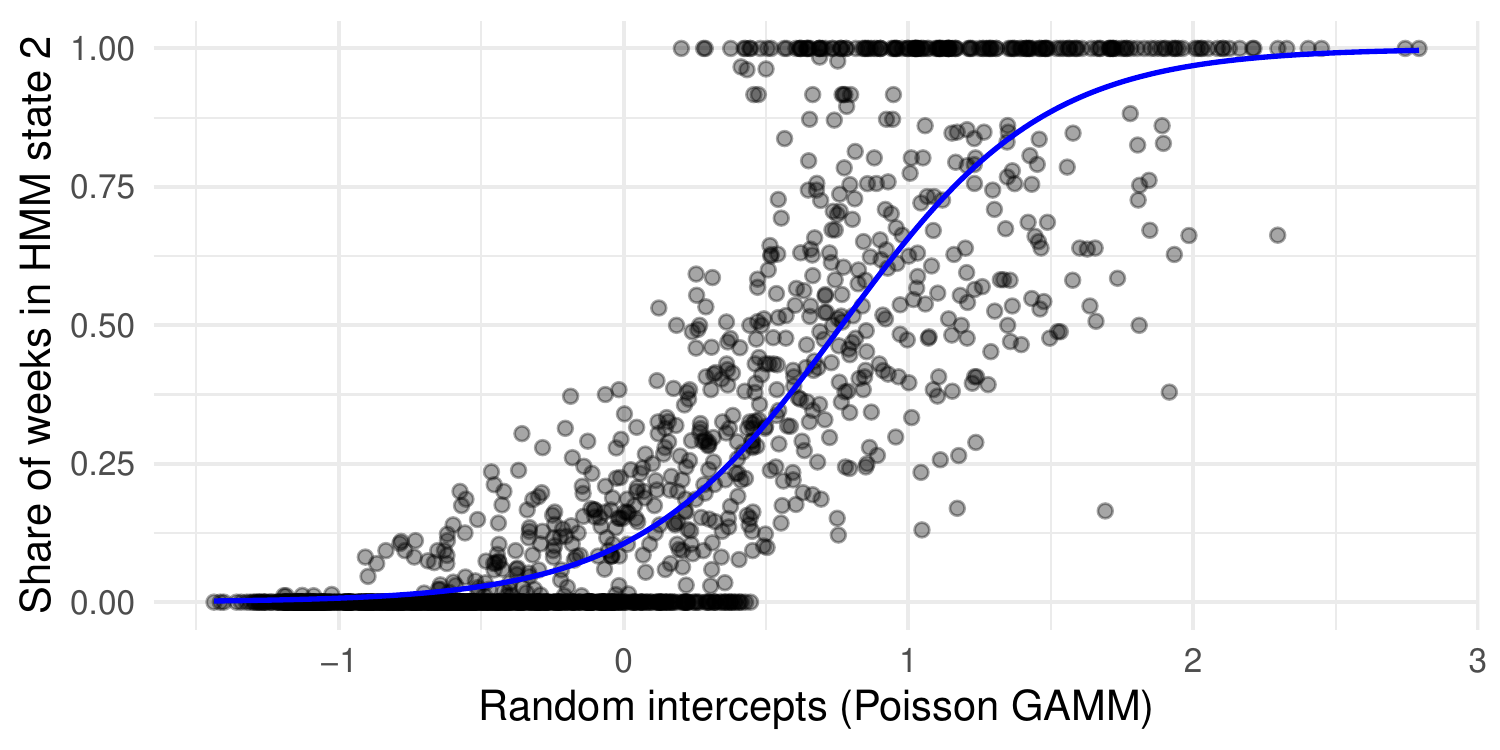}
  \caption{Borrower heterogeneity and regime occupancy. Each point shows a borrower's share of observed weeks classified in the high-intensity regime (State~2), plotted against the Poisson GAMM random intercept $u_i$. The curve is the fitted binomial-logit model $\text{logit}(p_i)=\alpha+\beta u_i$ with $\hat\beta=2.79$ (SE 0.02).}
  \label{fig:re_state2_share}
\end{figure}

For each borrower $i$, let $n_i$ denote the number of observed weeks and let $n_{i,2}$ denote the number of weeks classified in the high-intensity regime (State~2) under the most probable state sequence. We define the empirical State~2 occupancy share as $\hat p_i = n_{i,2}/n_i$. Figure~\ref{fig:re_state2_share} plots $\hat p_i$ against $u_i$ and shows a monotone relationship from near-zero occupancy at low $u_i$ to near-one occupancy at high $u_i$.

This relationship is also strong in rank terms, with a Spearman correlation of $\rho=0.83$ between $u_i$ and $\hat p_i$. To quantify the pattern in a model-based way, we fit a binomial-logit specification:
\[
n_{i,2}\sim\text{Binomial}(n_i,p_i),\qquad \text{logit}(p_i)=\alpha+\beta u_i.
\]
The estimated curve is shown in blue in the figure. The slope estimate is $\hat\beta=2.79$ (SE 0.02, $p<10^{-16}$), implying that a one-unit increase in $u_i$ multiplies the odds of being in State~2 in a given week by $\exp(\hat\beta)\approx 16$. This supports the interpretation that the HMM high-intensity regime corresponds closely to a high-propensity borrower subgroup captured by the GAMM random intercept. However, the HMM additionally captures within-borrower transitions into and out of this regime over time.

\subsection{Finite-mixture evidence for the random intercepts}\label{app:re_mixture}

To characterise the shape of the distribution of borrower random intercepts $\{u_i\}$, we fit Gaussian finite mixtures with $G\in\{1,\ldots,5\}$ components and select $G$ using BIC. Table~\ref{tab:re_bic} reports the BIC values. The best specification under this criteria is $G=4$. The estimated component weights for this specification are (0.12, 0.22, 0.31, 0.35), with means approximately (-1.09, -0.62, -0.09, 0.86), showing substantial heterogeneity and a pronounced high-intercept component (see Table~\ref{tab:re_mix_params}).

\begin{table}[ht!]
\centering
\begin{tabular}{lrrrrr}
\toprule
Components $G$ & 1 & 2 & 3 & 4 & 5 \\
\midrule
BIC & -4075.6 & -3819.8 & -3829.0 & \textbf{-3778.5} & -3798.1 \\
\bottomrule
\end{tabular}
\caption{BIC for Gaussian finite mixtures fitted to Poisson GAMM random intercepts $u_i$. Higher BIC is preferred.}
\label{tab:re_bic}
\end{table}

\begin{table}[ht!]
\centering
\begin{tabular}{lrrrr}
\toprule
Component & 1 & 2 & 3 & 4 \\
\midrule
Weight & 0.125 & 0.219 & 0.308 & 0.349 \\
Mean & -1.085 & -0.622 & -0.095 & 0.861 \\
Variance & 0.0149 & 0.0246 & 0.1679 & 0.3806 \\
\bottomrule
\end{tabular}
\caption{Estimated parameters of the selected $G=4$ Gaussian mixture for $u_i$.}
\label{tab:re_mix_params}
\end{table}

\subsection{Policy-based default vs random intercept heterogeneity}\label{app:re_policy_default}

\begin{figure}[ht!]
\centering
  \includegraphics[width=0.9\textwidth]{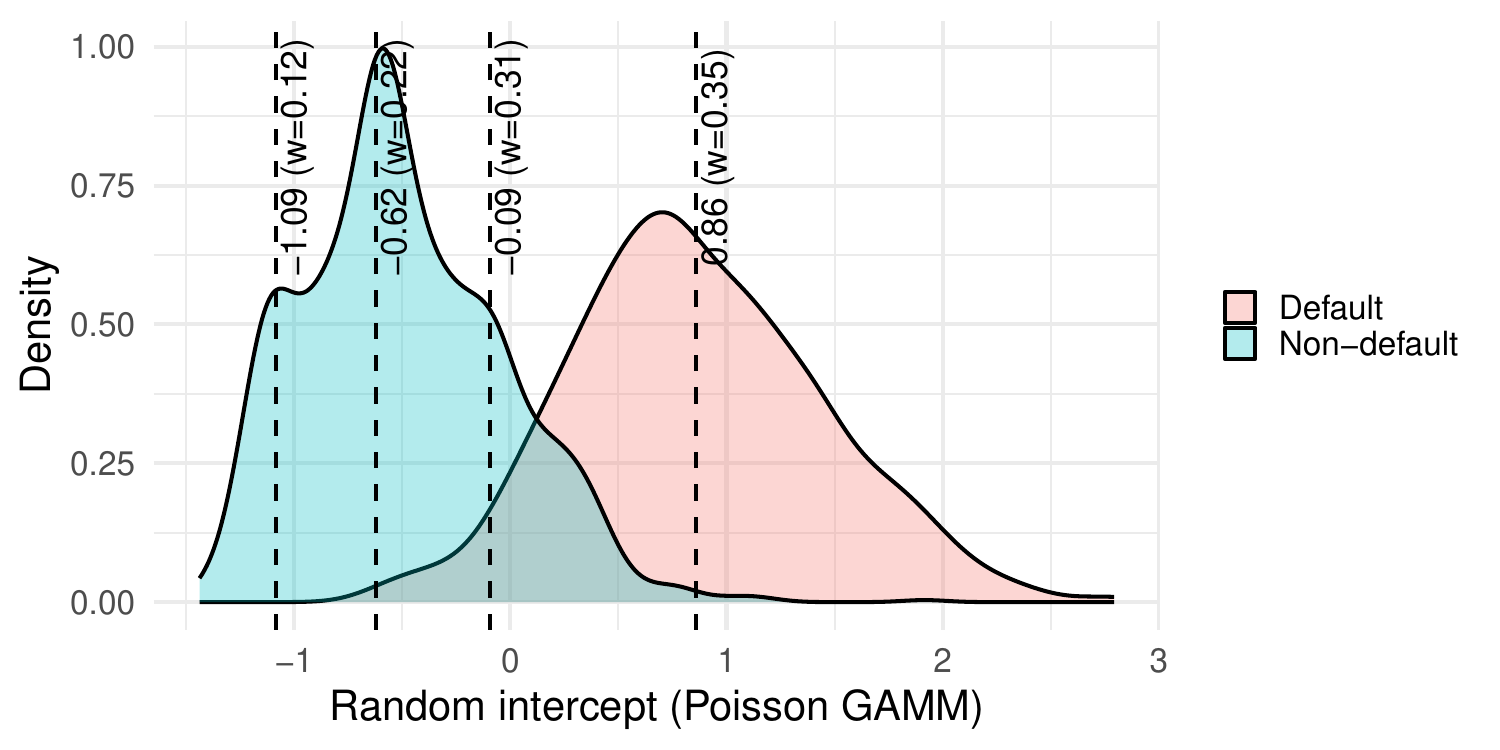}
  \caption{Distribution of borrower-specific random intercepts $u_i$ from the Poisson GAMM, stratified by inferred default status under the prolonged exposure policy. Dashed vertical lines show the means of the  four-component Gaussian mixture fitted to $\{u_i\}$ (labels report the mean and its mixture weight).}
  \label{fig:re_u_by_default}
\end{figure}

Using the policy measure defined in Section~\ref{subsec:emp_policy} (default defined as at least $H=12$ consecutive weeks in State~2), we examine how inferred default aligns with the heterogeneity captured by $u_i$. Figure~\ref{fig:re_u_by_default} shows that borrowers classified as default are concentrated in the high-$u_i$ region, whereas non-default borrowers concentrate at negative $u_i$.

Table~\ref{tab:re_default_comp} shows inferred default against mixture-component membership under the selected $G=4$ model. We observed that 81.5\% of inferred defaults are assigned to component~4, compared with 3.8\% of non-default borrowers. The default rate within component~4 is 92.5\%, compared with 9.9\% among borrowers in components 1--3. A chi-squared test strongly rejects independence ($p<10^{-16}$), indicating that the heterogeneity captured by the random intercept aligns closely with the policy.

\begin{table}[ht!]
\centering
\begin{tabular}{lrrrr}
\toprule
 & Component 1 & Component 2 & Component 3 & Component 4 \\
\midrule
Non-default & 221 & 430 & 360 & 40 \\
Default & 0 & 8 & 103 & 490 \\
\bottomrule
\end{tabular}
\caption{Inferred default status by mixture component (selected $G=4$ model).}
\label{tab:re_default_comp}
\end{table}

\end{document}